\documentclass[12pt]{iopart}

\usepackage{epsf}
\newcount\figno
\def\fig#1#2#3{
\par\begingroup\parindent=0pt\leftskip=1cm\rightskip=1cm\parindent=0pt
\baselineskip=11pt
\global\advance\figno by 1
\epsfxsize=#3
\centerline{\epsfbox{#2}}
\vskip 12pt
{\bf Figure \the\figno:} #1\par
\endgroup\par
}
\def\figlabel#1{\xdef#1{\the\figno
\mbox{ }}}
\def\encadremath#1{\vbox{\hrule\hbox{\vrule\kern8pt\vbox{\kern8pt
\hbox{$\displaystyle #1$}\kern8pt}
\kern8pt\vrule}\hrule}}

\def\drawbox#1#2{\hrule height#2pt
        \hbox{\vrule width#2pt height#1pt \kern#1pt
              \vrule width#2pt}
              \hrule height#2pt}
\def\Fund#1#2{\vcenter{\vbox{\drawbox{#1}{#2}}}}
\def\Asym#1#2{\vcenter{\vbox{\drawbox{#1}{#2}
              \kern-#2pt       
              \drawbox{#1}{#2}}}}
\def\fund{\Fund{6.5}{0.4}}
\def\asym{\Asym{6.5}{0.4}}
\def\href#1#2{#2}
\def\beq{\begin{equation}}
\def\eeq{\end{equation}}
\def\beqa{\begin{eqnarray}}
\def\eeqa{\end{eqnarray}}
\def\ov{\overline}

\newcommand{\be}{\begin{equation}}
\newcommand{\ee}{\end{equation}}
\newcommand{\bea}{\begin{eqnarray}}
\newcommand{\eea}{\end{eqnarray}}
\newcommand{\ena}{\end{eqnarray}}

\newcommand{\ba}{\begin{array}}
\newcommand{\ea}{\end{array}}
\newcommand{\beann}{\begin{eqnarray*}}
\newcommand{\eeann}{\end{eqnarray*}}

\newcommand{\Z}{{\mathbf Z}}
\newcommand{\R}{{\mathbf R}}
\newcommand{\C}{{\mathbf C}}
\newcommand{\IS}{{\mathbf S}}
\newcommand{\NN}{{\cal N}}
\newcommand{\id}{{\bf 1}}
\newcommand{\diag}{{\rm diag}}

\begin{document}
\rightline{MIT-CTP-3101}
\rightline{CERN-TH/2001-076}
\rightline{hep-th/0103177}

\vspace{1.0truecm}
\centerline{\bf \Large Orientifold dual for stuck NS5 branes}

\vspace{0.6truecm}
\centerline{\bf
Bo Feng, Yang-Hui He, Andreas Karch
}
\vspace{.2truecm}
{\em \centerline{Center for Theoretical Physics, MIT,
Cambridge, MA 02139, USA}}
{\em\centerline{yhe@grendel.mit.edu; fengb, karch@ctp.mit.edu}}
\vspace{.4truecm}
\centerline{\bf
Angel M. Uranga
}
\vspace{.2truecm}
{\em \centerline{TH Division, CERN, CH-1211 Geneve 23, Switzerland}}
{\em\centerline{Angel.Uranga@cern.ch}}

\vspace{0.5truecm}

\begin{abstract}
We establish T-duality between NS5 branes stuck on an orientifold 8-plane
in type I' and an orientifold construction in type IIB with D7 branes
intersecting at angles. Two applications are  discussed. For one we obtain
new brane constructions, realizing field theories with gauge group a
product of symplectic factors, giving rise to a large new class of
conformal ${\cal N}=1$ theories embedded in string theory. Second, by
studying a D2 brane probe in the type I' background, we get some
information on the still elusive (0,4) linear sigma model describing a
perturbative heterotic string on an ADE singularity.

\end{abstract}


\section{Introduction}

Supersymmetric gauge theories can be embedded
into string theory via intersecting branes and branes ending on branes,
following the pioneering work of \cite{hw}. This approach proved powerful
in predicting moduli spaces, global symmetries and dualities of the
theories engineered that way. Even these days, now that with the advent of
AdS/CFT duality we have more refined tools to actually study dynamics,
brane setups are still quite popular to help getting intuitive pictures.
This is facilitated by the fact that many brane setups can be related by
T-duality to branes moving on a singular geometry. In particular,
dualization into orbifold and orientifold backgrounds proves useful, since
this way one can employ perturbative string techniques to calculate and
derive the rules governing the brane setup. These T-dualities follow
from the duality between Kaluza-Klein monopoles and NS5
branes \cite{NSdual}. In \cite{ilka,hanzaf} it was first used to relate 6d
Hanany-Witten (HW) setups to D5 branes on orbifold singularities. Similar
relations were found for ${\cal N}=2$ theories in 4d \cite{fractional,pu},
D-brane probes of the conifold
\cite{uranga,dandm} and other CY singularities \cite{hanur,mina3}.

In this paper we will study T-duality relations for NS5 branes in type I'
theory. The building blocks are the NS5 branes along 012345, the defining
orientifold O8 planes and the corresponding D8 branes along 012345789. In
addition one may introduce D6 branes along 0123456, that is, stretching
along the direction 6 which is taken to be a compact interval. This is the
background geometry and it preserves 8 supercharges. On this background we
are going to consider various probes preserving 4 supercharges, in
particular a D2 brane along 016 or a D4 brane along 01237.

The low energy dynamics of the background is described by an ${\cal
N}=(1,0)$ gauge theory in 6 dimensions. There are two different phases,
shown in Fig. 1: the NS5 branes can either be free to move in pairs in the
bulk, their 6 position being the scalar in a tensor multiplet, or they can
be stuck on one of the O8 planes, with positions encoded in the scalars of
hypermultiplets. Taking the direction 6 compact, the former setup is
easily T-dualized along it into a type IIB orientifold on an 
ALE space\footnote{More precisely, on a Taub-NUT (TN) space. 
	The field theory data,
	however, depend only on the geometry near the centers, which can be
	approximated by an ALE geometry.} 
of the kind studied in
\cite{bs,bs2,gp,dp,gj,intblum}.
In this
T-dual orientifold picture the spectrum and interactions on the branes can
be reliably calculated. Due to the extra tensor multiplets, it is well
known that this setup does not correspond to a perturbative
compactification of ten-dimensional type I theory. In the S-dual heterotic
$SO(32)$ theory, we are describing a non-perturbative phase with small
instantons (T dual to the D6 branes) on an ALE singularity (T dual to the
NS5 branes). The requirement that the NS5 branes are moving in pairs
translates into the statement that some of the singularities are frozen
and cannot be blown up (for every NS5 brane pair we are always left with
at least an A$_1$ singularity).

The other phase is when the NS5 branes are stuck on the orientifold, as
studied in \cite{hanzaf2}. This of course can still be T-dualized in the
same way, however the T-dual is no longer a free orbifold conformal field
theory, so we no longer have a perturbative description. In this phase all
the tensors are frozen while all the NS5 branes can move independently,
hence it describes type I on the ALE space. However we are no longer
presented with a calculational tool to construct the spectrum on the probe
branes.

We will present another T-duality, along the 7 direction, that transforms
the setup with the stuck NS5 branes into a perturbative orientifold of IIB
with O7 planes and D7 branes intersecting at angles. This is a non-compact
version of the models introduced in \cite{ralph}. As an application and
check we will consider introducing D2 brane and D4 brane probes in this
background. The D4 brane realizes a new class of ${\cal N}=1$ theories in
four dimensions, including several conformal examples. The D2 brane
corresponds to a D1 brane probe in type I theory on an ALE singularity,
hence our construction provides some information on the so far elusive
(0,4) LSM \cite{wade} describing the heterotic string on an ALE space or,
once we include the D6 branes, on the ADHM constructions of $SO(32)$
instantons on an ALE space.

In the next section we will review the type I' background, describe the
T-duality along the 6 direction and review the problems associated with
stuck NS5 branes. In section 3 we introduce the orientifold construction
and give evidence that it is indeed the T-dual after dualizing along 7. In
the following two sections we then present as applications and checks the
theory on the D4 brane probe and the D2 brane probe.

\fig{Examples of the two different phases in the type I'
setup. Tensor multiplets correspond to motion on the compact 6 interval,
hyper-multiplets to motion in the transverse 789 space. The $m$
referred to the cosmological constant set up in the bulk.}
{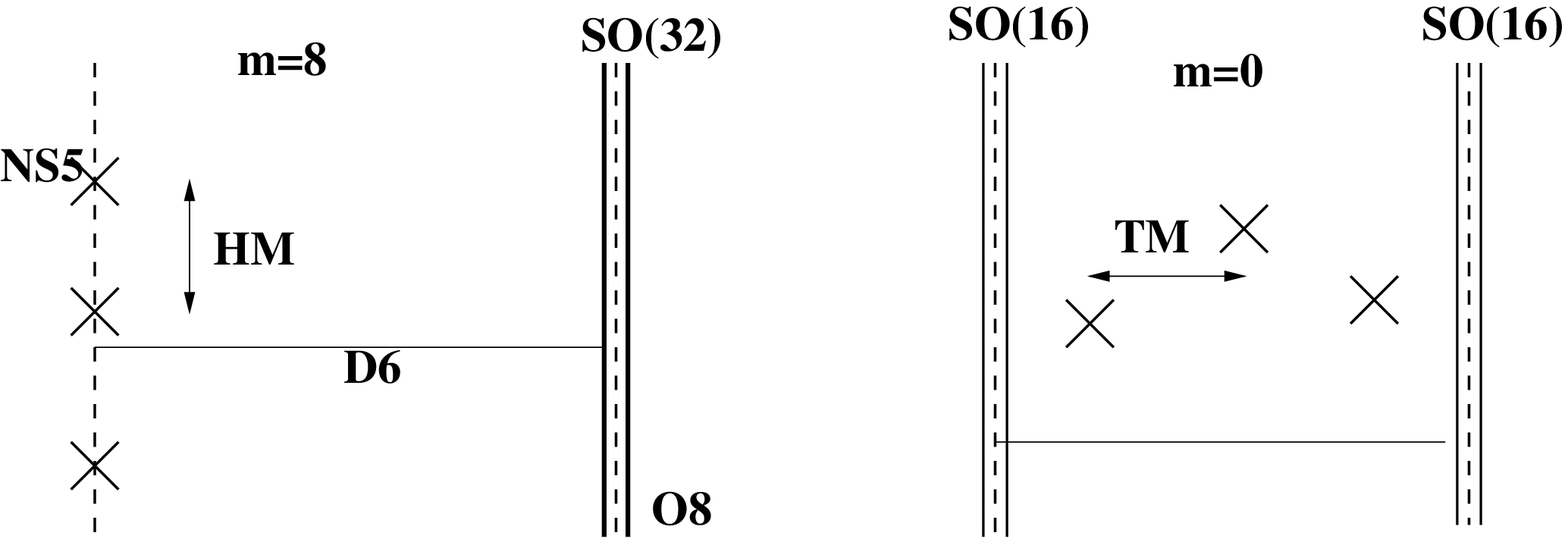}{9truecm}
\figlabel{\2phases}

\section{The type I' background and T-duality to type I}
\subsection{T-duality along the interval direction}

We start by reviewing the T-duality of type I' theory with O8 planes
(along 012345789), NS5 branes (along 012345), and D6 branes (spanning
0123456) along the compact 6 interval, as described in the introduction.
First let us discuss the case that corresponds to a perturbative
orientifold of type IIB on an ALE space, of the kind discussed in
\cite{bs,bs2,gp,dp,gj,intblum}. In the type I' dual the NS5 branes are
out in the bulk, half the hypers (corresponding to 789 positions) are
frozen and we have extra tensors from the 6 position of the NS5 branes
\cite{ilka,hanzaf}. Turning on Wilson lines on the IIB side
\footnote{More accurately, turning on asymptotically flat self-dual gauge
backgrounds in the Taub-NUT geometry. We denote the asymptotic connections
by `Wilson lines', and point out that in the ALE limit they correspond to
choices of D brane Chan-Paton factors for the orbifold group.} maps in
type I' to moving the D8 branes into the bulk. For a $\Z_k$ singularity
with odd $k$ on the IIB side, one NS5 brane is stuck on an O8 plane and
the others move in the bulk in pairs. For even $k$, we can have $\frac{k}{2}$ 
pairs or one NS5 brane stuck at each of the O8 planes and the rest moving in
pairs. The stuck NS5 branes correspond to monopoles living in the D8 gauge
group \cite{hanzaf2}. In the literature, e.g. \cite{intblum}, latter
models are often referred to as `without vector structure'.

Of course this IIB orientifold is not type I on the ALE space. In order to
obtain type I we have to mod out by world-sheet parity $\Omega$. The action
$\Omega$ reverses the sign of the NS-NS B-field, and hence is a symmetry of
type IIB string theory only if all the NS-NS B-fields are turned off.
Since in the free world-sheet orbifold CFT of type IIB on ALE space
the twisted sector NS-NS B-fields are non-zero, in this theory $\Omega$
can not be gauged. Instead we would have to orientifold the interacting
ALE conformal field theory at $B=0$.

In the perturbative orientifolds with non-zero B-fields of
\cite{bs,bs2,gp,intblum}
$\Omega$ is combined with a space-time action \cite{tensors} exchanging
oppositely twisted sectors. The resulting models involve tensor multiplets
and correspond to new phases of the heterotic string. The only example
without extra tensors (the orientifold of $\C^2/\Z_2$ in \cite{bs,bs2,gp})
corresponds to a bundle without vector structure \cite{blpssw}, which hence
also describes a compactification of the $SO(32)$ string with a
non-trivial gauge background turned on. In type I' the corresponding
configuration has one NS5 brane stuck at each O8 plane.

Unless we turn Wilson lines to the $SO(16) \times SO(16)$ point, additional
D6 branes are required in the bulk, due to charge conservation in the
background of the non-trivial cosmological constant \cite{ilka,hanzaf}. We
are always free to add an arbitrary equal number of D6-branes on each
interval, corresponding to adding small instantons (that is D5 branes) on
the IIB side.

\subsection{$SO(32)$ strings on a smooth ALE}

In order to study the heterotic string with a trivial bundle on an ALE, we
have to mod out IIB just by $\Omega$ (without geometric action) and study
the D-string in this background. As argued above, we would have to
orientifold the theory at $B=0$. Let us once more analyze the T-dual type
I' language. Since in this phase the twisted sector tensors are projected
out, the NS5 branes must be stuck on the orientifold planes, with their
positions within the O8 plane parametrized by scalars in the hypermultiplets. 
If we realize the $SO(32)$ by putting all D8 branes on (say) the right O8,
in order to have a trivial bundle we should locate all the NS5 branes at the 
left O8 plane. Having all the D8 branes on one side sets up a
cosmological constant in the bulk. A NS5 brane in such a background
cosmological constant would have to be connected with 8 D6 branes to the
right O8 plane. So without extra (fractional) small instantons present (D6
branes along 0123456), the bulk cosmological constant does not allow the
NS5 branes to wander off into the bulk. There is no phase transition
trading hypermultiplets  for tensors \cite{hanzaf2}. For a generic choice
of Wilson lines, the NS5 branes are stuck in a similar fashion on the less
occupied O8 plane. They can be interpreted as monopoles of the D8 brane
world-volume gauge theory. Only at the $SO(16) \times SO(16)$ point NS5
branes may move around freely, since the bulk cosmological constant
vanishes.

We will show that in order to describe the phase with stuck NS branes,
another T-duality can be employed. In the type I' language this is a
duality along the direction 7. This duality can be established when we
take the 6 direction to be non-compact, that is we study a single O8 with the
stuck NS5 branes, while compactifying the 7 direction. So this
is not really the T-dual of the situation we want to study, with 7
non-compact and 6 compact. However when interested in the gauge theory on
a probe, all the interesting dynamics are determined locally by the
interplay of probe branes and orientifold and NS5 branes. The new
T-duality gives us a calculational tool to describe a single O8 plane with
an arbitrary number of D8 branes and stuck NS5 branes, and probes on this
background.

\section{A new T-duality into a orientifold with branes at angles}
\subsection{A non-compact orientifold with branes at angles}

According to \cite{NSdual}, $k$ NS5 branes on a circle are T-dual to an ALF 
space with a $\C^2/\Z_k$ singularity at the origin. Positions of the NS5
brane in the transverse space map to the 3 blowup parameters associated with 
each of the $(k-1)$ homologically non-trivial 2-spheres, and positions
along the compact direction map to the NSNS 2-form field fluxes through
the spheres. Note that having the NS5 branes stuck on the O8 freezes one
of the 3 blowup modes (the 6 position). Instead we have a 10 position on
the M-theory circle, which corresponds to a RR 2-form flux in the IIB
orientifold. 

In our configuration of Figure \2phases, 
considering the direction 7 compact and T-dualizing
along it, the T-dual is an $k$-center Taub-NUT (TN) space, with the circle
fiber parametrized by $7'$, the T-dual of the 7 direction, and the base
parametrized by 689. Our purpose is to identify the geometry of the T-dual
of the type I' O8/D8 system. The original O8 planes and D8 branes were
wrapped in the 7 direction, hence they should correspond to D7 branes not
wrapped on the circle fiber of TN. In a suitable complex structure, the TN
can be described by the hypersurface in $\C^3$
\begin{equation}
UV=Z^k, 
\end{equation}
and the circle corresponds to the $U(1)$ orbit $(U,V)\to(e^{i\lambda} U,
e^{-i\lambda}V)$, with real $\lambda$ ranging from $0$ to $2\pi$. 
Near the core of the TN space, the geometry is locally that of a
$\C^2/\Z_k$ singularity, with the generator $\Theta$ of $\Z_k$ acting as
\begin{equation}  \Theta:\cases{ z_1 \to e^{2\pi i/k}\, z_1 \cr
              z_2 \to e^{-2\pi i/k}\, z_2,\cr}
\end{equation}
where one can take e.g. $z_1 := x^6 + ix^8$ and $z_2 := x_7 + i x_9$.
These coordinates are related to the above ones by $U=z_1^k$, $V=z_2^k$,
$Z=z_1 z_2$.

\vspace{.2cm}
\fig{On the covering space the orbifold action acts on
the branes/planes by rotation. Including all the mirror images
under the orbifold action, we are required to include branes
intersecting at angles.}
{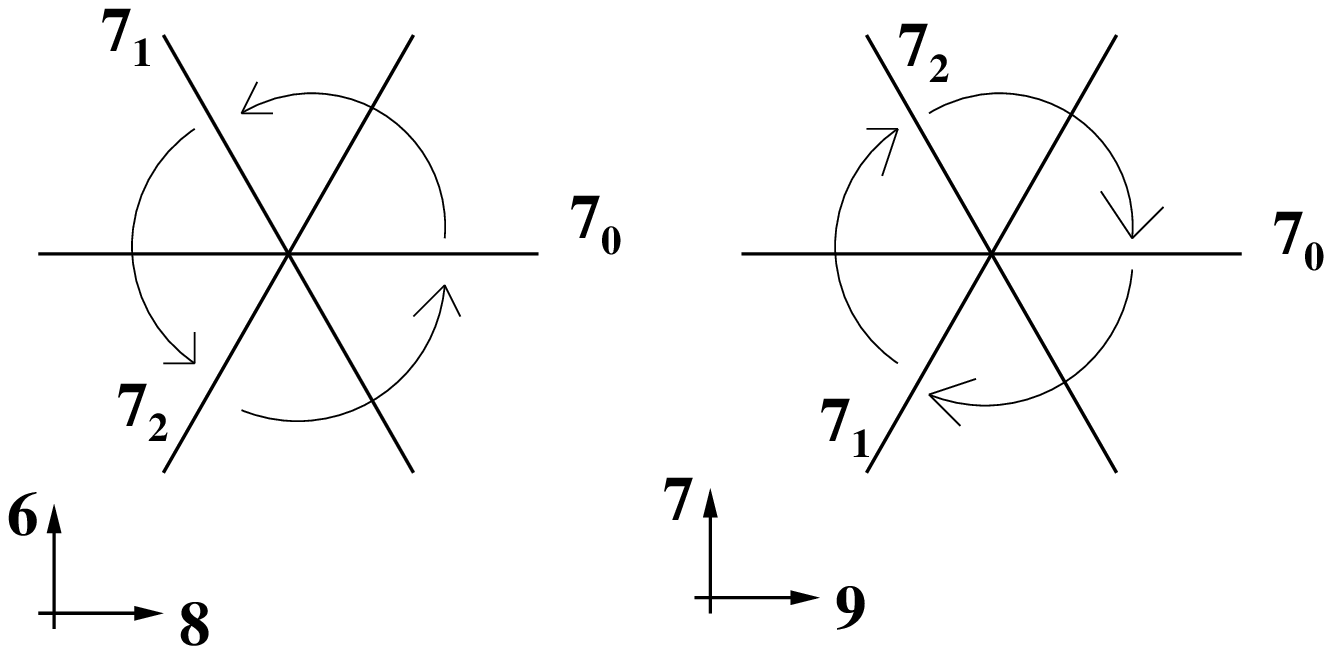}{5truecm}
\figlabel{\atangle}

So in order to T-dualize the O8 plane we are looking for an orientifold
action in the Taub-NUT geometry, with sets of fixed points (O planes) not
wrapping the $\IS^1$ fiber. An action with the correct properties is given
by $\Omega R (-)^{F_L}$, where $\Omega$ is world-sheet parity, and $R$ is
the spacetime action 
\begin{equation}  R:\cases{ z_1\to {\ov z}_1 \cr
              z_2\to {\ov z}_2 \cr}
\label{orientifold}
\end{equation}
The fixed set of $R$ is $z_1={\ov z}_1$, $z_2={\ov z}_2$, which is a
special Lagrangian cycle. Hence the corresponding O7 plane preserves the
correct number of supersymmetries. Notice that the full orientifold group
contains different elements $\Omega R \Theta^a$, whose sets of fixed
points lead to a set of $k$ O7 planes at angles, as shown in Figure
\atangle.
Specifically, the curves wrapped by the O7 planes are given by
\beqa
z_1=e^{-2\pi i \frac {a}{k}} \ov{z}_1 \quad ; \quad 
z_2=e^{2\pi i \frac {a}{k}} \ov{z}_2.
\label{o7curves}
\eeqa
with $a=0,\ldots, k-1$.
These orbifold and orientifold actions have been considered in \cite{ralph}, 
in the context of compact toroidal orbifolds.

Clearly, the T-duals of the D8 branes correspond to $\Z_k$ invariant sets
of D7 branes at angles wrapped on curves of the type described, see Figure
\atangle. To see this, for example, if we start with a D7 brane
wrapped on $z_1={\ov z}_1,z_2={\ov z}_2$, after the action of $\Theta^b$
we get a D7 brane wrapped on $e^{i\frac{2\pi b}{k}} z_1 =e^{-i\frac{2\pi
b}{k}} {\ov z_1}$ and $e^{-i\frac{2\pi b}{k}} z_2 =e^{i\frac{2\pi b}{k}}
{\ov z_1}$. Rewriting this we have $z_1 =e^{-i\frac{4\pi b}{k}}{\ov z_1}$
and  $z_2 =e^{i\frac{4\pi b}{k}}{\ov z_1}$, which is a curve of the kind
in (\ref{o7curves}) for $a=2b$. Furthermore, this calculation shows that
there is a difference between the $k$ odd case and $k$ even case. For
odd $k$ a $\Z_k$ invariant configuration of D7 branes is given by $k$ sets
of D7 branes wrapped on the curves above. For even $k$, however, $\Z_k$
does not relate curves with even and odd $a$. Hence it is possible to
construct $\Z_k$ invariant combinations of D7 branes with only $P=k/2$
sets, related by $\Z_P$, and the orbifold action contains an additional
$\Z_2$ acting within each stack. Even though there is no inconsistency in
such possibilities, we will be interested in configurations with local
charge cancellation (see Section 3.4). Such configurations involve $k$ sets
of D7 branes on the $k$ curves above, namely two $\Z_k$ invariant sets of 
D7 branes, wrapped on the even and odd $a$ curves, respectively. 

A similar construction can be made for orbifold CY 3-folds. In this
case the branes will wrap special Lagrangian 3-cycles. Compact versions of
such models have appeared in \cite{ralph4d, fhs}. Non-compact orbifolds
with branes at angles (in the absence of orientifold projection) have been
considered in \cite{gaberdiel}.

\subsection{The closed string spectrum and the worldvolume theory
on the 7-branes}

\subsubsection{The closed string spectrum}

Besides the usual matter from the untwisted sector, the closed string
twisted sectors contain, before the orientifold projection, $k-1$ hyper
and tensor multiplets of 6d $(1,0)$ susy. The orientifold projection above
maps each twisted sector to itself, and can be seen to project out the
tensor multiplets, leading to $k-1$ hypermultiplets, in agreement with the
result in the type I' construction. This result generalizes to arbitrary
$\Z_k$ the closed string spectrum for crystallographic twists in
\cite{ralph}. 

An alternative derivation is to follow the analysis of \cite{angelangle},
which deals with the related orientifold action $\Omega R'$ with 
$R':(z_1',z_2')\to (z_2',-z_1')$, and leads to $k-1$ tensor multiplets
and no hypermultiplets. This is similar to our action $R$ if we rewrite
$R$ by expressing the same ALE in a different preferred complex structure,
by defining $z_1'=z_1+\bar{z_2}$, $z_2'=z_2+\bar{z_1}$, where we have
$R:(z_1',z_2') \to (z_2',z_1')$. This differs from $R'$ in just one sign,
which can be seen to flip the orientifold action in the twisted sectors to
yield $k-1$ hypermultiplets and no tensors.

\subsubsection{The worldvolume theory on the D7 branes}

In order to calculate the worldvolume theory we have to specify the
$\Z_k$ action on the D7 brane indices. Starting with the odd $k$ case,
and labeling the $k$ stacks of $n$ D7-branes by a Chan-Paton index
$a=1,\ldots,k$, the action of the generator $\Theta$ of $\Z_k$ is to map
the $a^{th}$ to the $(a+1)^{th}$ stack. Hence we have
\beqa
{\small \gamma_{\Theta,7} = {\pmatrix{ 
& \id_n & & &  \cr
& & \id_n & & \cr                         
& & & \ldots & \cr
& & & & \id_n \cr
\id_n & & & & \cr }}}
\label{gamone}
\eeqa
which we write as ($\gamma_{\Theta,7})_{ab}=\delta_{b,a+1} \id_{n}$.
Notice that upon diagonalization, this matrix is equivalent to the more
familiar
\begin{equation} 
\label{theta7}
\gamma_{\Theta,7}=\diag(\id_n, \omega \id_n, \ldots, \omega^{k-1} \id_n)
\end{equation} 
where $\omega=e^{2\pi i/k}$. However, working on the basis in
(\ref{gamone})
is more convenient. 

The orientifold projection is represented by the matrices
\beqa
\gamma_{\Omega R,7} & = & \diag (A_n,\ldots,A_n) 
\eeqa
with 
\be
A_n=\id_N
\ee 
or 
\be
A_n={\pmatrix{0 & \id_{n/2} \cr -\id_{n/2} & 0}}
\ee
corresponding to the choice of O8$^+$ or O8$^-$ plane on the T-dual side
respectively.

For even $k=2P$, we consider configurations with two $\Z_k$ invariant set
(a total of $k$ stacks), which we treat independently. Each contains $P$
stacks of $n$ D7-branes, labeled by $a=1,\ldots, P$. The $\Z_k$ action is
represented by the $P\times P$ block matrix
\beqa
{\hspace{-2cm}
{\small \gamma_{\Theta,7} = {\pmatrix{ 
& M_n & & &  \cr
& & M_n & & \cr                         
& & & \ldots & \cr
& & & & M_n \cr
M_n & & & & \cr }}} \quad {\rm with} \;\;
M_n=\diag(e^{\pi i/k} \id_{n/2}, e^{-\pi i/k} \id_{n/2})
\label{gamtwo}
}
\eeqa
namely $(\gamma_{\Theta,7})_{ab}=\delta_{b,a+1} M_n$. The orientifold
action is given by
\beqa
\gamma_{\Omega R,7} & = & \diag(A_n,\ldots, A_n) 
\eeqa
with 
\be
A_n={\pmatrix{0 & i\id_{n/2} \cr i\id_{n/2} & 0}} \ee
or
\be
A_n={\pmatrix{0 & \id_{n/2} \cr -\id_{n/2} & 0}}
\ee
corresponding to the O8$^+$ or O8$^-$ plane on the T-dual side.

\medskip

Let us discuss the spectrum after the orbifold projection, but before the
orientifold projection. The results are shown in the first half of table
\ref{specs}. Recall that we start with a $\Z_k$ action and have to
distinguish the case of even and odd $k$. For odd $k$ we have $k$ sets of
$n$ branes, which we denote as D7$_a$-branes. For even $k=2P$ we have two 
$\Z_k$-unrelated families of $P$ sets.

The matter content consists of an 8D piece and some matter localized
at the 6D intersection. For the purposes of discussion, let us momentarily
pretend that the D7 branes are somehow `compactified' and phrase the
spectrum in terms of $D=6$ ${\cal N}=1$ SUSY. Namely we discuss the
structure of the zero modes of the 8d fields in the bulk of the D7 branes.
In the 7$_a$7$_a$ sectors, we start with a gauge group $U(n)^k$ and
adjoint superpartners. For odd $k$, the $\Z_k$ simply maps one set of
branes to the next, and so reduces the group to just one $U(n)$ and the
matter to just one adjoint hypermultiplet. For even $k=2P$, the
$\Z_k$ projection on the original $U(n)^P\times U(n)^P$ vector plus
adjoint hypermultiplets can be regarded as acting in two steps. First, the
$\Z_P$ projection reduces to $U(n)\times U(n)$. Next, the remaining
$\Z_2$ maps each stack to itself, and projects the vector multiplets to
$[U(n/2)\times U(n/2)]^2$, and the adjoint hypers to hypers in two
copies of the $(n/2,n/2;1,1)+(1,1;n/2,n/2)$. These spectra are easily
obtained using the projection by the matrices (\ref{gamone}),
(\ref{gamtwo}) on the original 7$_a$7$_a$ spectrum, namely 6d $\NN=2$
vector multiplets of $U(n)^k$.

Let us turn to the 7$_a$7$_b$ sectors, for $a\neq b$. From each such sector
we obtain one half-hypermultiplet (the other half comes from the 7$_b$7$_a$
sector, which we count as a different one for the moment) in the 
bifundamental of $U(n)_a \times U(n)_b$. For odd $k$ we get $k(k-1)/2$
hypermultiplets in such representations, which are projected by $\Z_k$ to
$(k-1)/2$ hypermultiplets in the adjoint of the surviving $U(n)$. For even 
$k$, open strings within each $\Z_k$ invariant set give $P(P-1)/2$ hypers
in bifundamental representations. The $\Z_P$ projection would
leave $(P-1)/2$ full hypers in the adjoint of each $U(n)$, which are
projected down to $(P-1)$ full hypers in the $(n/2,n/2)$ of
each $U(n/2)\times U(n/2)$ by the additional $\Z_2$ projection. Open
strings stretched between the two $\Z_k$ invariant sets give $P^2$
bifundamental hypers, which are projected down to $P$ hypers in the
bifundamental of $U(n)^2$ by the $\Z_P$ projection. The additional $\Z_2$
projection leaves $P$ hypers in the $(n/2,1;n/2,1)+(1,n/2;1,n/2)$ of
$U(n/2)^2\times U(n/2)^2$.

\medskip

Let us now consider introducing the orientifold projection, associated for
example to an O8$^-$ plane in the type I' T-dual. The result is shown in
the second half of table \ref{specs}. Let us again first look at the
7$_a$7$_a$ sector. For odd $k$ the orientifold projects the single $U(n)$
down to $SO(n)$, and the matter to a hypermultiplet in the adjoint. For
even $k$, within each $\Z_k$ invariant set the orientifold projection
relates the two $U(n/2)$ factors. For each we obtain one $U(n/2)$ gauge
group, and two hypers in two-index antisymmetric representation.

In the 7$_a$7$_b$ sector for $a \neq b$, imposing the orientifold
projections for odd $k$ projects the $(k-1)/2$ adjoint hypermultiplets
of $U(n)$ to adjoints of $SO(n)$. For even $k=2P$, orientifolding of
open strings within each $\Z_k$ invariant set leads to $(P-1)$ full hypers
in the two-index antisymmetric representation of $U(n/2)$. The
projection on the spectrum of open strings stretched between different
invariant sets yields $P$ hypermultiplets in the bifundamental of
$U(n/2)^2$.

\begin{table}[htb] \footnotesize
\renewcommand{\arraystretch}{1.25}
\begin{center}
\begin{tabular}{|c|c|c|c|c|}
\hline 
6d (1,0) multiplet &       & Vector & $7_a7_a$ hyper & $7_a7_b$ hyper \\
\hline\hline 
Orbifold & $k$ odd & $U(n)$ & Adj. & $\frac{k-1}{2}$ Adj. \\
\cline{2-5} 
         & $k$ even & $U(n/2)^2$ & $2(\fund,\fund;1,1)$&
$(P-1)\,\, [(\fund,\fund;1,1) + (1,1;\fund,\fund)]$ \\
 & &$\times U(n/2)^2$ 
&$+2(1,1;\fund,\fund)$ & $P\,\, [(\fund,1;\fund,1) + (1,\fund;1,\fund)] $
\\
\hline\hline
Orientifold & $k$ odd & $SO(n)$ & $\asym$ & $\frac{k-1}{2}\,\, \asym$ \\
\cline{2-5} 
         & $k$ even & $U(n/2)^2$ & $2\,\, (\,\asym\,;1)+ 2\,\,
(1;\,\asym\,)$ & $(P-1)\,\, [\,(\,\asym\,;1)+(1;\,\asym\,)\,]$ \\
& & & & $P\,\, (\fund;\fund)$ \\
\hline \end{tabular}
\end{center}
\caption{\small Spectrum on D7-branes at angles in orbifold and
orientifold singularities.
\label{specs} }
\end{table}

\vspace{.2cm}
\fig{
The 8d modulus for $k=3$.
}
{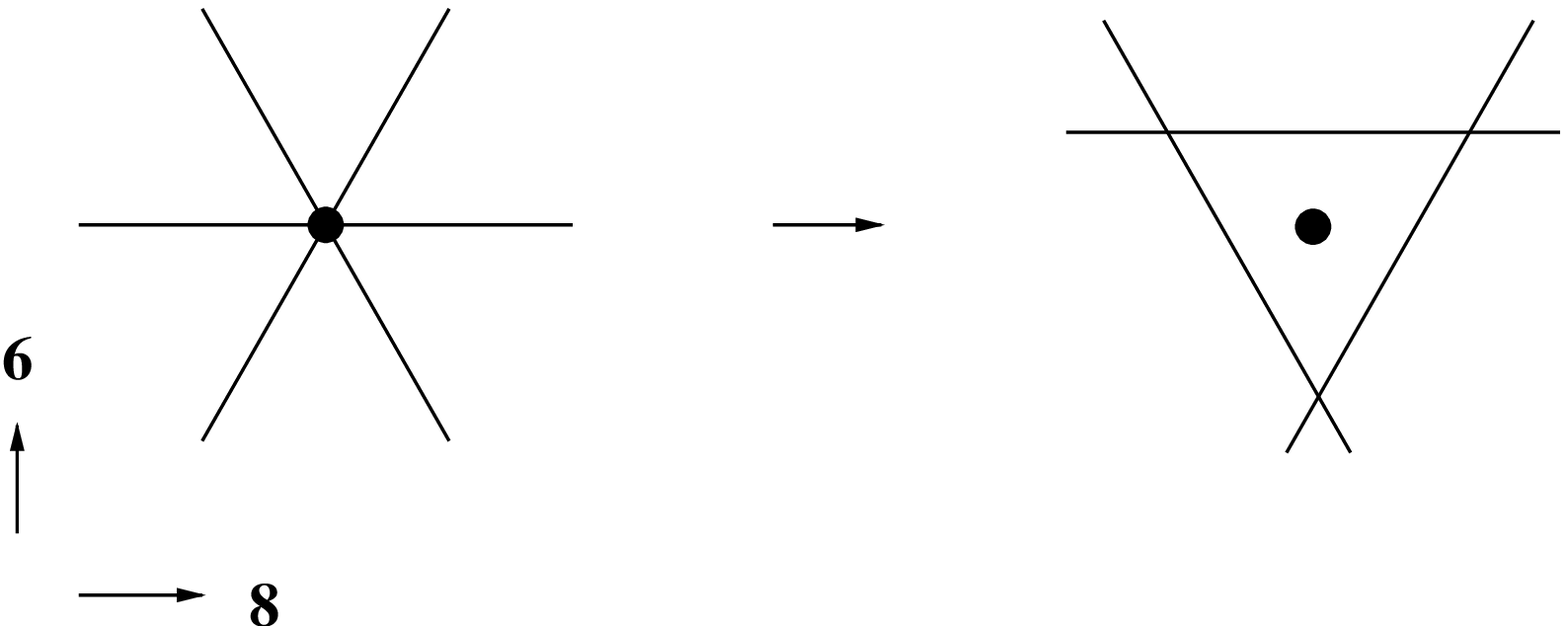}{6truecm}
\figlabel{\ngon}

\vspace{.2cm}
\fig{
The 8d modulus for $k=5$. The two kinds of intersections lead to two
hypermultiplets.
}
{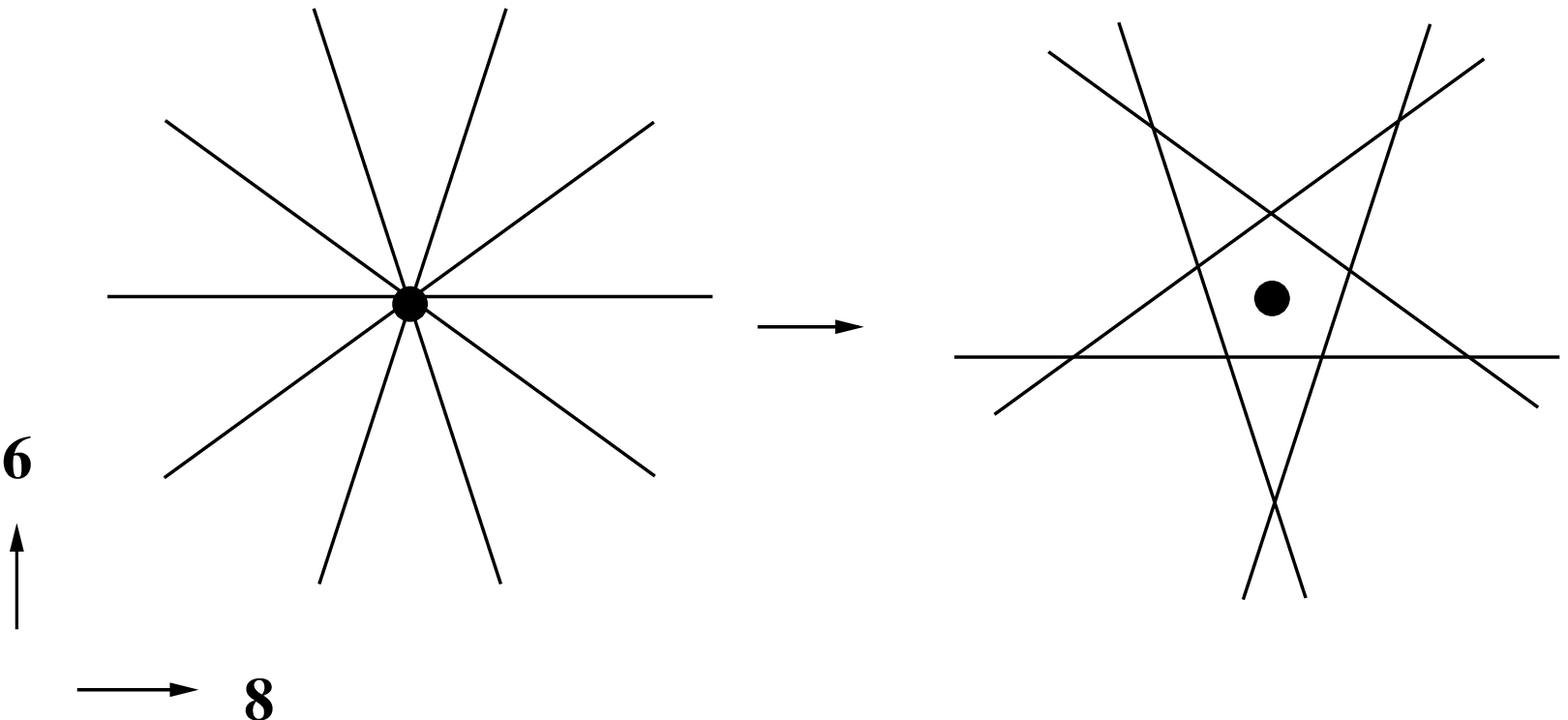}{6truecm}
\figlabel{\5gon}

For a single brane the interpretation of the scalar moduli in these 
multiplets is as follows: The scalars in the D7 brane bulk correspond to
the motion of the D7 brane away from the fixed points. Due to the orbifold
symmetry all the mirror images have to move as well, so that afterwards
the branes form a regular $k$-gon, as displayed for $k=3$ in Fig \ngon
and for $k=5$ in Fig \5gon. 
In this configuration every brane still intersects every other brane. At
each intersection there lives one of the hypermultiplets. For instance,
the case $k=5$ in Fig \5gon contains two kinds of intersections,
associated to two hypermultiplets in the 7$_a$7$_b$ sectors. Turning on
the vevs for such hypermultiplets corresponds to deforming the
intersecting 7-brane configuration into a smooth curve, as in
\cite{mina3}. All intersections that are mapped into each other under the
orbifold symmetry of course have to be turned on simultaneously, giving a
nice geometric interpretation of the above counting of multiplets.

Let us conclude by mentioning that the six-dimensional chiral fermions
localized at the intersections lead to an anomaly, which is nevertheless
canceled by an anomaly inflow mechanism from the bulk of the D7 branes
\cite{ghm,scruse,scruse2}

\subsection{Closed string - open string duality}

As already noted in early references \cite{cargese,ps,bs,bs2}, an
important restriction on open string configurations is the requirement
that open and closed strings couple in a consistent fashion. By this we
mean that the annulus, M\"{o}bius strip and Klein bottle amplitudes,
computed in the open channel as 1-loop, should admit a consistent
description in the closed channel as tree-level propagation between
boundary and/or crosscap states.

This requirement has been studied in setups with branes at angles in
\cite{ralph}, where strong consistency conditions were derived.
The case addressed in \cite{ralph} was on compact orbifold models and
the above requirements imposed non-trivial restrictions on the choice of
compactification lattices and orientifold actions on them. Our case is
non-compact, and there is no such freedom as choosing a compactification
lattice, hence one might worry about consistency of the models. In this
section we show that the resulting models satisfy the requirements
of open-closed duality.

Let us briefly go through the general procedure for the cylinder
amplitude, which is enough to illustrate the point. We also restrict to
odd $k$ for simplicity. The cylinder amplitude in the open string channel
is obtained by tracing over the open string spectrum and
performing the orbifold and GSO projections. The result in our present
context, for open strings stretching between the $a^{th}$ and
$(a+r)^{th}$ stack of D7-branes, is easily obtained following the
indications in \cite{ralph}
\begin{equation}
{\cal A}_{\,r} = c(1-1) \int_{0}^{\infty} \frac{dt}{t^4} \frac{n^2}{2}
{{\vartheta{\left[\begin{array}{c} 0 \\ 1/2
\end{array}\right]}^2 
\vartheta\left[\begin{array}{c} r/k \\ 1/2 \end{array}\right]
\vartheta\left[\begin{array}{c} -r/k \\ 1/2 \end{array}\right] }
\over{\eta^6 \;\,
\vartheta\left[\begin{array}{c} -1/2+r/k \\ 1/2 \end{array}\right]
\vartheta\left[\begin{array}{c} 1/2-r/k \\ 1/2 \end{array}\right] }}
\end{equation}
where the theta and eta functions have argument $q=e^{-2\pi t}$. The constant 
$c$ encodes numerical factors irrelevant to our analysis. Also, for $r=0$
one should include momentum states, and some theta functions actually
become eta functions, but we ignore this point to avoid cluttering.

In \cite{ralph} the above amplitude was multiplied by a numerical factor
corresponding to the intersection number of the D7 brane stacks. These
factors played a crucial role in satisfying open-closed duality. Here we
show that, even though our models do not contain such factors (there is
only one intersection), a consistent amplitude is obtained in the tree
channel.

Going to the tree channel by a modular transformation, $t=\frac 1{2l}$,
the resulting amplitude is
\begin{equation}
{{\tilde{\cal A}}}_{\,r} = c(1-1) \int_{0}^{\infty} dl\, n^2
{{{\tilde \vartheta}\left[\begin{array}{c} 1/2 \\ 0 \end{array}\right]^2 
{\tilde \vartheta}\left[\begin{array}{c} 1/2 \\ r/k \end{array}\right]
{\tilde \vartheta}\left[\begin{array}{c} 1/2 \\ -r/k \end{array}\right] }
\over{{\tilde \eta}^6 \;\,
{\tilde \vartheta}\left[\begin{array}{c} 1/2 \\ -1/2+r/k \end{array}\right]
{\tilde \vartheta}\left[\begin{array}{c} 1/2 \\ 1/2-r/k \end{array}\right] }}
\label{treeone}
\end{equation}
where the modular functions ${\tilde \vartheta}$ and ${\tilde \eta}$ are
the usual $\vartheta$ and $\eta$ but with argument ${\tilde q}=e^{-4\pi
l}$. This should admit the interpretation of tree-level closed string
exchange between D7 brane boundary states, schematically,
\be
{\tilde{\cal A}_{r}} \sim \langle D7_a| q^{L_0} {\ov q}^{\ov L_0}
|D7_{a+r}\rangle.
\label{treetwo}
\ee
Clearly, the result is independent of $a$ since only the relative angle
between D7 branes is relevant. Since the rotated state $|D7_{a+
r}\rangle$
is simply obtained by applying $\Theta^{r}$ to $|D7_a \rangle$, we may
write
\beqa
{\tilde{\cal A}_r} \sim \langle D7| q^{L_0} {\ov q}^{\ov L_0} 
\Theta^r |D7\rangle.
\eeqa
with bra and ket representing states of parallel D7 branes. The amplitude
(\ref{treeone}) is easily seen to have the right structure: the numerator
(resp. denominator) represents summing over the fermionic (resp. bosonic)
oscillator states excited by the D7 brane boundary state, with the
shifted lower characteristics in the theta functions corresponding to
$\Theta^r$ insertions. However, there is an important numerical factor
that should also match. This factor appears because the theta functions of
(\ref{treeone}) have upper characteristic $1/2$, and have a product
expansion
\beqa
{\tilde \vartheta}\left[\begin{array}{c} 1/2\\ 1/2+\phi\end{array}\right]
= 2 \sin(\pi \phi) [\;\eta(\tau) {\tilde q}^{\frac 1{12}}
\prod_{n=1}^\infty (1+{\tilde q}^n e^{2\pi i\phi}) (1+{\tilde q}^n
e^{-2\pi \phi}) \;].
\eeqa
For theta functions in the numerator the factor $2 \sin(\pi \phi)$ is
expected, since it is associated to the trace of $\Theta^r$ over fermion
zero modes. For theta functions in the denominator such a factor does not
arise in the trace over bosonic oscillators, and hence our tree-level result 
obtained from the loop channel by duality appears with an additional
factor of $1/(4\sin^2 \pi r/k)$. As mentioned above, in \cite{ralph} the
original loop amplitude had an additional multiplicity from multiple
intersections, which (along with some numerical factors from the structure
of the lattice) canceled the problematic factor, leading to correct
tree channel amplitudes.

Our models are nevertheless consistent, because there is indeed an
explanation for this factor in the tree amplitude. It arises from tracing
over the momentum states excited by the D7 brane boundary state. Their
contribution can be evaluated in analogy with a similar trace computed in
section 4.3 of \cite{gj}. Exchange in the tree channel involves momentum
states, which form a continuum in the non-compact limit. In \cite{gj} the
trace of $\Theta^r$ over a continuum of momentum states in the four
non-compact dimensions of $\C^2/\Z_k$ yielded $1/(4\sin^2(\pi r/k))^2$. In 
our case, momentum states excited by the D7 brane boundary state
correspond to only two directions, hence give only `half' of the
contribution, $1/(4\sin^2(\pi r/k))$, explaining that the factor implicit
in (\ref{treeone}) is actually correct.

More specifically, the trace of $\Theta^r$ over a continuum of momentum
modes is, in position space 
\be
\int dx_6\, dx_8\, \langle x_6, x_8 | \Theta^r | x_6 x_8 \rangle,
\label{pos_int}
\ee
which, defining $z_1=x_6+ix_7$, $z_2=x_8+ix_9$, is equal to
\bea
\int d^2z_1\, d^2z_2\, \delta(x_7)\,\delta(x_9)\, 
\langle z_1, z_2|e^{2\pi ir/k}z_1,e^{-2\pi ir/k} z_2 \rangle  \\
\nonumber
= \int d^2z_1\, d^2z_2\, \delta(x_7)\,\delta(x_9)\,
\delta^{(2)}((1-e^{2\pi i r/k})z_1)\, \delta^{(2)}((1-e^{-2\pi ir/k})z_2)
\\
\nonumber
= 1/(4\sin^2(\pi r/k)).
\eea
It is interesting to compare our result with the large volume limit of a
compact example, of the type studied in \cite{ralph}. In the compact case,
there are precisely $4\sin^2(\pi r/k)$ $\Theta^r$-fixed points per complex
plane, which is integer for crystallographic $\Z_k$ ($k=2,3,4,6$). Hence
there is a cancellation of contributions in the integral from each fixed
point with their total number, whereby giving no net multiplicity.
In our non-compact case of $\C^2/\Z_k$, the unique fixed points gives the
factor $1/(4\sin^2(\pi r/k))$ in the tree channel amplitude, consistently
with the open string loop result.

\subsection{Tadpoles}

Consistency of the configuration would require cancellation of RR tadpoles
which are not volume suppressed, i.e.  RR charges whose flux cannot
escape to infinity. As usual, untwisted tadpoles are not required to
cancel, since they are volume suppressed and we work on non-compact
setups. On the other hand, twisted RR tadpoles arising from disks
associated to D7-branes, or from crosscaps associated to the orientifold
projection would, if non-zero, lead to an inconsistency, since the sources
for the corresponding charge fill all non-twisted non-compact directions,
leading to no volume suppression.

Fortunately there are no such twisted tadpoles. Following \cite{ralph}
one can see that in either the Klein bottle, Moebius strip, or cylinder,
only untwisted modes propagate in the tree channel amplitude (as can be
seen in (\ref{treeone}) for the cylinder). By factorization, this means
that the branes and the orientifold planes do not carry charges under
twisted modes. Hence the models are consistent without any constraint on
the brane content of the theory.

For future application, it is however interesting to study configurations
where untwisted tadpoles associated to D7 branes do cancel. We emphasize again
that this is not required for consistency. However, it leads to the
interesting property that the closed string fields have flat profiles in
the resulting configuration. When some D brane probe is introduced in the
theory, like D3 branes in Section 4, varying profiles correspond to
running coupling constants (see e.g. \cite{fractional, leroz}), and flat
profiles correspond to theories with no running, namely finite theories.

The value of the untwisted tadpoles associated to the D7 branes can be
extracted from the analysis of \cite{ralph}, and gives the expected
answer. For either odd or even $k$, each of the $k$ O7 planes has charge
$-8$ (counted in D7 brane charge units, in the covering space), as
usual. Hence cancellation of untwisted tadpoles is achieved for $n=8$,
namely eight D7 branes on top of the each O7 plane. 

Note that our construction seemingly leads to a puzzle. In the original HW
type I' picture we have one O8 plane along 012345789 and several NS branes
along 012345, with 7 a compact coordinate. T-dualizing along 7 we obtain a
Taub-NUT space, with $k$ coincident centers, so that locally we have
$\C^2/\Z_k$. One would expect that after T duality, the O8 plane would map
into two O7 planes sitting at opposites sides of the $\IS^1$ fiber in the
TN space. Our proposed T-dual is however (centering on odd $k$ for
simplicity) just one O7 plane along $z_1=\bar{z}_1$, $z_2=\bar{z}_2$, and
its orbifold images. Another related puzzle is that in the original type
I' local charge cancellation is achieved for 16 D8 branes overlapping with
the O8 plane, whereas in our type IIB picture it is achieved by 8 D7
branes overlapping the O7 planes.

These puzzles are solved because the T-dual description we are using
is valid only in the near center region of Taub-NUT space, whereas the
intuition about what the T-dual should be is good far from it, where the
geometry splits as a (twisted) product of $\R^3$ and $\IS^1$. In order to
extrapolate our description of the T-dual orientifold planes to the region
far from the TN core, and compare with the intuitive expectations,
the simplest way is to identify the $\IS^1$ orbit in $\C^2/\Z_k$ and count
the number of intersections with the O7 plane. The $\IS^1$ is the $U(1)$
orbit $(e^{i\lambda} U, e^{-i\lambda} V)$, for $\lambda$ from 0 to
2$\pi$, in $UV=Z^k$. The O7 plane wraps the curve $U={\ov U}$, $V={\ov
V}$, which can be parametrized by taking real $U$, $V$. This 2-cycle
intersects the $U(1)$ orbit at two opposite points. Hence by continuously
deforming the $\IS^1$ fiber to the region far from the TN core, we
learn that our single O7 plane looks like two O7 planes sitting at
opposite points in $\IS^1$ in the asymptotic geometry. Similarly, the
single set of 8 D7 brane in the near center region looks in the asymptotic
region like 16 D7 branes, in two sets at opposite sides of $\IS^1$.
Hence our construction works just as required to reproduce the intuitive
T-dual picture.

\section{New brane constructions of 4d ${\cal N}=1$ theories}

As a first application of our type IIB orientifold construction, let us
study a D3 brane probe on the orientifold geometry. In type I' this is a
D4 brane along 01237, embedded in the O8 plane and stretched in between
the stuck NS5 branes.
\vspace{.2cm}
\fig{
The D4 brane probe in type I'.
}
{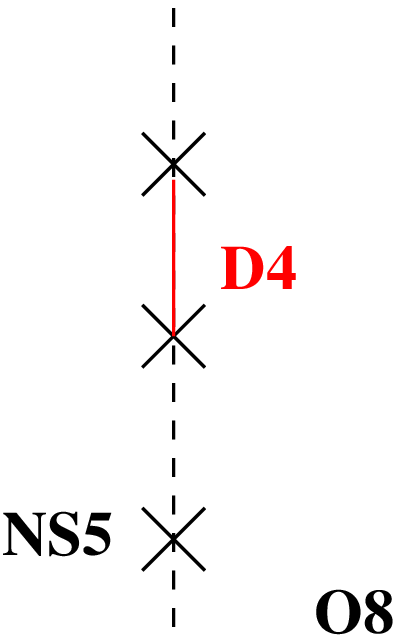}{3truecm}
\figlabel{\d4}
\subsection{The brane and field theory calculations}

As pointed out in \cite{amiandofer}, in addition to the standard branes
for realizing ${\cal N}=1$ theories in 4 dimensions (see \cite{review}),
one may introduce one more component, a D8 brane in which the D4 brane is
embedded. In our type I' picture this is realized when we introduce a D4
brane probe oriented along 01237. In the T-dual picture this corresponds
to introducing a D3 brane probe along 0123. We will analyze the matter
content expected from field theory and HW construction considerations, and
then compare this with the actual calculation on the IIB side.

Let us analyze first the gauge theory on a D4 embedded in a D8 brane.
As a second step we will introduce the O8 plane. Both are well known
SUSY gauge theories with 8 supercharges. The final step is to study the
gauge theories of D4 branes ending on NS5 branes, with the whole setup
embedded inside the D8/O8 system. 

A stack of $N$ D4 branes inside $n$ D8 branes on the compact 7 circle has
an ${\cal N}=2$ SUSY $SU(N)$ gauge theory on the 4d non-compact piece of
their worldvolume. The matter content is that of the ${\cal N}=4$ SUSY
theory on the D4 brane (a ${\cal N}=2$ vector and adjoint hypermultiplet) 
plus $n$ additional hypermultiplets in the fundamental representation. The
two adjoint scalars in the vector multiplet parameterize motion away from
the D8 branes, and the four adjoint scalars in the hypermultiplet
correspond to motion within the D8 brane. Turning on vevs for the $n$
extra hypers resolves the D4 branes into instantons in the D8 brane gauge
group. Adding an O8$^-$ plane, one obtains an ${\cal N}=2$ SUSY $USp(N)$
gauge theory with a hypermultiplet in the antisymmetric tensor
representation and $n/2$ additional fundamental hypermultiplets, with
$SO(n)$ global symmetry. Again the two scalars in the adjoint in
the vector multiplet parameterize motion away from the O8 plane, while the
four scalars in the antisymmetric tensor hypermultiplet parameterize
motions within the O8 plane, and the fundamentals resolve the D4 branes
into instantons. Similarly, introducing instead an O8$^+$ plane, we can
achieve an $SO(N)$ gauge theory with a symmetric tensor hypermultiplet and
$n/2$ fundamentals, with $USp(n)$ global symmetry.

Now consider $N$ D4 branes with $k$ NS5 branes embedded in $n$ D8 branes, 
first in the absence of orientifold projection. The resulting theory has
${\cal N}=1$ SUSY in 4d. The gauge group is $SU(N)^k$. Each gauge factor
has $n$ fundamental and $n$ antifundametal chiral multiplets $Q_i$,
$\tilde{Q}_i$ ($i=1,\ldots, k$) from 4-8 strings. In addition we have the
standard bifundamental chiral multiplets $F_{i, i+1}$, $\tilde{F}_{i,
i+1}$ from strings stretching across the NS5 branes. The adjoint
hypermultiplet which corresponded to motions in 789 and to the Wilson line
along 6 is eliminated by the NS brane boundary condition. However, there
remains the adjoint chiral multiplet $X_i$ from the $\NN=2$ vector
multiplet, parameterizing the 45 motion. The 1-loop beta function of a
given factor is proportional to
\beqa
3\mu_{adj}-\mu_{matter}= 3(2N)-2N-2 \cdot N\cdot (1+1) - n \cdot (1+1) =-2n,
\eeqa
leading to an asymptotically non-free theory due to the extra D8 brane matter.
The superpotential is
\beqa
W & = &  \sum_i \left[ \, F_{i, i+1} X_i \tilde{F}_{i, i+1}+
\tilde{F}_{i-1, i} X_i F_{i-1, i}+  \right. \nonumber \\
&& \quad \left. + Q_i \tilde{F}_{i, i+1} \tilde{Q}_{i+1}+
\tilde{Q}_i F_{i, i+1} Q_{i+1}\right].
\eeqa
The first two terms are the relics of the $\NN=2$ system formed by the NS
and D4 branes. The last two terms are allowed by gauge invariance, and
should be included in order to break the global symmetry from the D8
branes from $SU(n)^k$ down to $SU(n)$.

Including the O8$^-$ plane, we obtain a $USp(N)^k$ gauge group with an
antisymmetric  chiral multiplet $A_i$, one set of bifundamentals $F_{i,i+1}$, 
and $n$ extra fundamentals $Q_i$ in each group factor. The global symmetry
is $SO(n)$, and the superpotential reads as above:
\beqa
W= \sum_i \left[ \, F_{i,i+1} A_i F_{i,i+1}+ F_{i-1,i} A_i F_{i-1,i}
+ Q_i F_{i,i+1} Q_{i+1}\,\right].
\label{supo}
\eeqa
For $n=8$ we should obtain a finite theory. Indeed the 1-loop beta
function is
\beqa
3\mu_{adj}-\mu_{matter}= 3(N+2)-(N-2)-2N-n=8-n,
\eeqa
which vanishes for $n=8$. In order to check whether the theory is actually
finite to all orders we perform an analysis following Leigh and Strassler 
\cite{leighstrassler}. Namely we show that the requirement that all
(exact) beta functions vanish actually leads to linearly dependent
equations, generically leading to lines of solutions instead of isolated
solutions in coupling space. Since in our scenario all superpotential
terms are cubic and the 1-loop beta vanishes, this line will pass
through the origin of coupling space, i.e. weak coupling. Hence, along the
line not only beta functions, but also the anomalous dimensions, will
vanish and the theory is indeed finite.

The beta functions for the gauge coupling and the three terms in the
superpotential are proportional to
\begin{eqnarray*}
\beta_{g_i} &\sim& N \gamma_{F_{i,i+1}} + N \gamma_{F_{i-1,i}}+
(N-2) \gamma_{A_i}+ 8 \gamma_{Q_i} \\
\beta_{W^1_i} & \sim & 2 \gamma_{F_{i,i+1}} + \gamma_{A_i} \\
\beta_{W^2_i} & \sim & 2 \gamma_{F_{i,i-1}} + \gamma_{A_i} \\
\beta_{W^3_i} & \sim & \gamma_{F_{i,i+1}} +\gamma_{Q_i} +\gamma_{Q_{i+1}}
\end{eqnarray*}
Defining $\gamma_F = \sum_i \gamma_{F_{i,i+1}}$, 
$\gamma_A = \sum_i A_i$, $\beta_g = \sum_i \beta_{g_i}$ we see that these
equations can be summed to obtain
\begin{eqnarray*}
\beta_{g} &\sim& 2N \gamma_{F} + 
(N-2) \gamma_{A}+ 8 \gamma_{Q} \\
\beta_{W^1}= \beta_{W^2} & \sim & 2 \gamma_{F} + \gamma_{A} \\
\beta_{W^3} & \sim & \gamma_{F} +2 \gamma_{Q}
\end{eqnarray*}
Since $W^1$ and $W^2$ are derived from the same $\NN=2$ relic we should put the
corresponding couplings equal, as reflected in above. The equations above
easily lead to the linear relation
$$\beta_g = 4 \beta_{W^3} + (N-2) \beta_{W^1},$$
which shows the existence of the desired line of RG fixed points.

\subsection{The spectrum from the orientifold calculation}

Let us reproduce the above results by performing the calculation in the
T-dual type IIB orientifold construction. Locating $Nk$ D3 branes at the
origin in $\C^2$, strings stretching among D3 branes lead to an $\NN=1$
vector multiplet $V$ with group $U(Nk)$, and three adjoint chiral
multiplets $X^1$, $X^2$, $X^3$, associated e.g. to the positions in
$z_1'=z_1+i{\ov z}_2$, $z_2'={\ov z}_1+iz_2$ and $z_3=x^4+ix^5$,
respectively. Strings stretched between D3 branes and each D7$_a$ brane
stack lead to chiral multiplets $H^{1,2}$ in the corresponding
bifundamental representations.

In order to reproduce a finite theory, there should not exist D3 brane
twisted tadpoles. This requires the action of $\Z_k$ on D3 branes to be
represented by a matrix 
\begin{equation}
\label{theta3}
\gamma_{\Theta,3}=\diag(\id_N, \omega \id_N, \ldots, \omega^{k-1} \id_{N}).
\end{equation}
with $\omega=e^{2\pi i/k}$. Representations other than the regular are
consistent, but lead to non-finite theories.

The orientifold action maps each eigenspace of $\gamma_{\Theta,3}$ to
itself. It is possible to show, following an analysis similar to section
2.2 of \cite{angelangle}, 
that the symmetry of $\gamma_{\Omega R,3}$ is equal
in all subspaces. Hence, the projection corresponding to, for example 
an O8$^-$ plane in the T-dual is
\begin{equation}
\label{omega3}
{\hspace{-1cm}
\gamma_{\Omega R,3}=\diag(M_N,M_N,\ldots ,M_N),~~~{\rm with}~~
M=\left(  \begin{array}{ll}  0   & \id_{\frac{N}{2}}  \\
-\id_{\frac{N}{2}} & 0 \end{array}  \right) =-M^{-1}
}
\end{equation}

The orbifold projection reads
\begin{equation}
\label{orb_pro}
\begin{array}{lllllll}
V & = &  \gamma_{\Theta,3}\, V\, \gamma_{\Theta,3}^{-1} & & & & \\
X^1 & = & \omega\, \gamma_{\Theta,3}\, X^1\, \gamma_{\Theta,3}^{-1} &
\quad ; \quad &
H^{1} & = & \gamma_{\Theta,3}\, H^{1}\, \gamma_{\Theta,7}^{-1}\\
X^2 & = & \omega^{-1}\, \gamma_{\Theta,3}\, X^2\, \gamma_{\Theta,3}^{-1} &
\quad ; \quad & 
H^{2} & = & \gamma_{\Theta,7}\, H^{2}\, \gamma_{\Theta,3}^{-1}\\
X^3 & = & \gamma_{\Theta,3}\, X^3\, \gamma_{\Theta,3}^{-1} & & & & 
\end{array}
\end{equation}
The 3-3 spectrum after the orbifold projection contains $\NN=1$ vector
multiplets of $U(N)^k$ (as usual, the $U(1)$ factors are expected to
disappear by the T-dual of the bending mechanism in \cite{wittenlift}),
chiral multiplets $F_{i,i+1}$, ${\tilde F}_{i,i+1}$ in bifundamental
representations, and $X_i$ in the adjoint. 

In the $37+73$ sector, for odd $k$, the orbifold projection simply 
identifies the sets of D7$_a$ branes and splits the D3 brane group. So
after the orbifold projection we get chiral multiplets $Q_i$, ${\tilde
Q}_i$ in representations $(N_i,n)$. For even $k=2P$, the $\Z_P$ projection
leads to a D3 brane group $U(2N)^P$, and leads to two sets of chiral
multiplets in the $(2N_i;n,1)+(2N_i;1,n)$. The additional $\Z_2$
projections breaks the D3 group down to $U(N)^k$, and the D7 group to 
$U(n/2)^2\times U(n/2)^2$, and leads to two sets of chiral multiplets in
the $(N_{2i};n/2,1;1,1)+ (N_{2i+1};1,n/2;1,1)$ $+(N_{2i};1,1;n/2,1)+
(N_{2i+1};1,1;1,n/2)$.

The orientifold projection is
\begin{equation}
\label{ori_pro}
\begin{array}{lllllll}
V & = & - \gamma_{\Omega R,3}\, V^{T}\,\gamma_{\Omega R,3}^{-1} & & & & \\
X^1 & = & \gamma_{\Omega R,3}\, (X^2)^T\, \gamma_{\Omega R,3}^{-1} & \quad 
;  \quad &
H^1 & = & \gamma_{\Omega R,3}\, (H^2)^T\, \gamma_{\Omega R, 7}^{-1} \\
X^2 & = & \gamma_{\Omega R,3}\, (X^1)^T\, \gamma_{\Omega R,3}^{-1} & \quad
; \quad &
H^2 & = & \gamma_{\Omega R,7}\, (H^1)^T\, \gamma_{\Omega R, 3}^{-1} \\
X^3 & = & \gamma_{\Omega R,3} (X^3)^{T}\gamma_{\Omega R,3}^{-1} & & & & 
\end{array}
\end{equation}

The 3-3 spectrum is as follows: there are vector multiplets of 
$USp(n)^k$, chiral multiplets $A_i$ in the antisymmetric
representation, and one set of chiral multiplets $F_{i,i+1}$ in
bifundamentals. 

In the mixed sector, the orientifold projection relates the 37 and 73
sectors. For odd $k$ we obtain one set of chiral multiplets in
representations $(N_i,n)$ of the $i^{th}$ $USp(N)$ factor in the D3 branes
and the D7 brane $SO(n)$ group. The case $n=8$ leads to 8 fundamental
flavours for each symplectic factor, and corresponds to a finite theory,
as anticipated from the untwisted tadpole computation in section 3.4.
For even $k$ we obtain chiral multiplets in the representations
$(N_i;n/2;1)+(N_i;1,n/2)$ of the $i^{th}$ $USp(N)$ D3 brane factor, and
the D7 brane $U(n/2)^2$. Again, for $n=8$ we obtain 8 chiral multiplets in
the fundamental of $USp(N_i)$, as required for finiteness.

Hence the corresponding spectra agree with those obtained in the HW brane
construction. Superpotential interactions are also easily seen to agree
with (\ref{supo}). This provides a check and a nice application of our
proposed T duality for stuck NS branes.

\section{The heterotic string on an ADE singularity}

\subsection{The problem}

In the linear sigma model (LSM) approach of \cite{phases},
instead of directly writing down the conformally
invariant non-linear sigma model describing the propagation of a string
in a given background, one starts with a 2d gauge theory, which in
the UV is just a free theory. Under the renormalization group flow
the couplings evolve and settle to their conformal values. As shown
in \cite{phases,edeva}, as long as we ensure the existence of an anomaly free
R-symmetry, we can expect a non-trivial CFT in the IR. Otherwise
the model will just flow to a massive and hence free theory.

D1 branes naturally provide us with 2d gauge theories. At strong coupling,
that is in the deep IR, these D-brane gauge theories flow to the
non-linear sigma model of the fundamental string in an S-dual background.
In this way the IIB string can be constructed as the ${\cal N}=(8,8)$ SUSY
theory of the D1 brane in IIB, the (0,8) theory of the D1 string in type I
gives us the heterotic string \cite{polchinski} and the Coulomb branch of
the D1 D5 system describes fundamental strings propagating in the
torsional metric set up by NS5 branes \cite{comments}.

In the same spirit we would like to study the LSM living on the
worldvolume of a D1 brane in type I on an ALE space. In the T-dual type I'
picture, the linear sigma model lives on a D2 brane along 016 stretched on
the interval between two O8 planes. The linear sigma model on the D2
brane has (0,4) supersymmetry. In the phase with the NS5 branes out in the
bulk the worldvolume theory is easily determined by using the T-duality
along 6. The gauge group consists of a (4,4) supersymmetric sector, which
is a quiver like theory with bifundamental matter and gauge group
$$ SO(1) \times U(1) \times U(1) \times \ldots \times U(1) \times SO(1)$$
for a $\Z_{2P}$ singularity and a vector bundle with vector structure, and
$$ U(1) \times U(1) \times U(1) \times \ldots \times U(1) \times U(1)$$
with extra `symmetric tensors', i.e. singlets, in the two factors at the
end of the chain for a $\Z_{2P}$ singularity and a vector bundle with
vector structure, and
$$ SO(1) \times U(1) \times U(1) \times \ldots \times U(1) \times U(1)$$
with an extra singlet in the last $U(1)$ for $\Z_{2P+1}$.
In addition we couple to the (0,8) matter sector from the D8 branes,
breaking the SUSY down to (0,4). These matter fields encode the
gauge bundle.

Heterotic string theory in this phase has marginal operators corresponding
to blowing up $P$ of the 2-spheres, to NS-NS fluxes through these spheres,
and to turning on the vevs for $P$ tensor multiplets ($P-2$ for the
case with two stuck NS5 branes). In the linear sigma model these
correspond to FI terms, theta angles and ratios of gauge couplings. While
the former two are marginal couplings of the CFT we flow to in the IR,
the role of the latter is not clear. In the analog (4,4) situation, type
IIB D1 string probe on ALE spaces, the ratio of gauge couplings correspond
to RR 2-form fluxes through the spheres of the ALE space. It can be shown
\cite{michaofer} that the ratios are actually irrelevant on the Higgs
branch theory which flows to the NLSM describing the IIB string on the ALE
space, and that they are only marginal on the Coulomb branch. This is in
agreement with the fact that CFT is mostly insensitive to RR potential
backgrounds, which only modify the one-point function on the disk. It
would be interesting to study if the situation changes in the (0,4)
context \footnote{We are indebted to Ofer Aharony and Mike Douglas for
illuminating discussions on this point.}.

Instead, we would like to understand the LSM on the D2 brane in the other
phase, that is with stuck NS branes leading to hypermultiplet moduli,
and no tensor multiplets. This provides the LSM for heterotic string
theory on the ALE space. Inclusion of D6 branes then yields a LSM
model realizing an ADHM construction for SO instantons on the ALE space.

\subsection{Using the new T-duality}

As has been discussed above, one can obtain some information of the D1
brane probe in type I theory on an ALE space by considering a D2 brane
probe (along 017) in the type I' model with NS branes stuck at one of the
O8 planes. We consider the situation with the dimension 7 compactified on
a circle, and study the system after T-dualizing along that direction. We
also momentarily consider the dimension 6 to be non-compact. As
shown in section 3, the set of NS branes transforms into a k-centered TN
space (different from the original one in type I), and the O8 plane maps
into O7 planes corresponding to orientifold actions of type 
(\ref{orientifold}).

The location of the TN centers in 89 correspond to the original 89
locations of the NS branes. We are interested in the case of coinciding
centers, and the geometry near the TN core is that of $\C^2/\Z_k$. On the
other hand, positions of NS branes in 7 correspond to NS-NS B fluxes on
the collapsed two-cycles, hence the perturbative orbifold description,
where all B-fields are equal, forces us to consider the NS branes equally
distributed on the 7 circle. This implies that in the original type I
theory we are studying the theory at $B=0$, but with non-zero blow-ups in
the direction 7. Hence, the information we can obtain from the T-dual
picture actually is associated to the LSM for the case of blown-up ALE, as
we will see below.

\fig{The 2-cycles in the Taub-NUT geometry, for D7/O7 and for D3 branes.}
{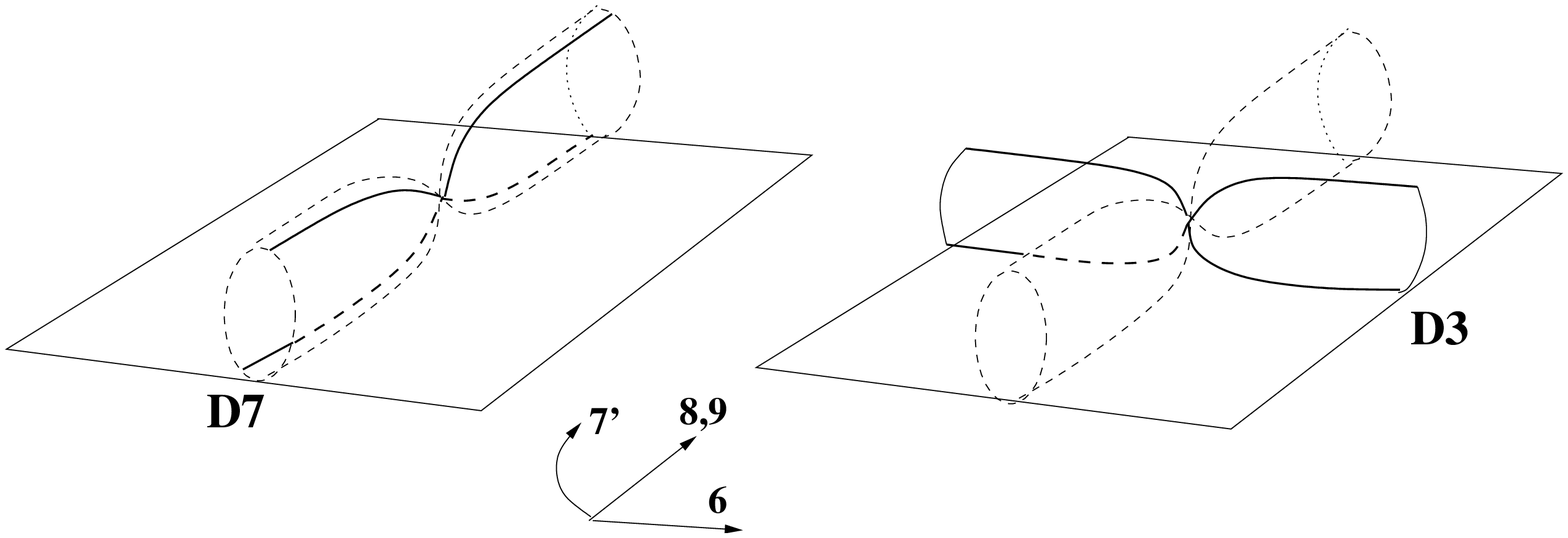}{11truecm}
\figlabel{\taubnut}

Let us discuss the T duality on a D2 brane ending on a NS brane.
The type I' D2 brane is not wrapped on the 7 circle, hence we expect it to
map to a D3 brane wrapped on a two-cycle in the TN geometry. The T-duality
of this object is very similar to that of D6 branes ending on NS branes,
analyzed in \cite{pru}, except that we are using a different preferred
complex structure. In our case, the corresponding 2-cycle is a special
Lagrangian cycle wrapped along the $\IS^1$ fiber in TN. In fact, in the
situation with the dimension 6 is non-compact, `half' D2 branes ending
on a NS brane, extending along either positive or negative values of
$x^6$, map to different 2-cycles. Moreover, since the orientifold action
flips the sign of $x^6$, it exchanges both kinds of half branes, hence in
the T-dual picture $\Omega R$ should exchange the two 2-cycles. 

This determines that the two 2-cycles are described, in the covering space
of the $\C^2/\Z_k$ orbifold, as
\beqa
z_1=i {\ov z}_2, \; \; {\ov z}_1=-i z_2 \quad {\rm and} \quad
z_1=-i {\ov z}_2,\;\; {\ov z}_1=i z_2
\label{twoslags}
\eeqa
Notice that they indeed are wrapped on the $U(1)$ orbit of the background
geometry. Also notice that they are related by an $SU(2)$ rotation to the
2-cycle associated to D7/O7 (e.g. $z_1={\ov z}_1$, $z_2={\ov z}_2$) hence
they preserve the correct number of supersymmetries. Finally, since the
above curves are invariant under the orbifold, we need not include
$\Z_k$ image D3 branes.
The
2-cycles associated to the D7/O7 system, and to the D3 branes are depicted
in Fig.\taubnut

When the direction 6 is considered compact, the T-dual geometry is an `ALG
space', with two asymptotically compact directions, which can be
constructed as an infinite (periodic in 6) array of k-centered TN spaces.
In this situations, the two above 2-cycles are in fact joined smoothly in
the $x^6$ region opposite the TN core. This is in analogy with the way the
two half D2 branes are smoothly joined on the side of the 6 circle
opposite to the location of the NS brane in the type I' picture.

Let us compute the spectrum and low energy effective action on such D3
brane probe. It is useful to first consider the situation without
orientifold projection, and also with non-compact 6. This exercise is
analogous to that performed in \cite{pru} for sets of half D6-branes.

First, notice that D3 branes wrapped on the 2-cycles (\ref{twoslags}),
subsequently denoted D3 and D3' branes, are fixed by the orbifold
action, hence we have the possibility of specifying a non-trivial action
of the Chan-Paton indices. The general choice would be
\beqa
\gamma_{\theta,3} & = & \diag(\id_{n_0},..., \omega^{k-1} \id_{n_{k-1}})
\eeqa
and a similar expression for D3' branes. Geometrically, these matrices
specify flat connections on the D brane bundles on the asymptotic region
of $\C^2/\Z_k$, or, in the context of TN (rather than ALE) geometry,
asymptotic Wilson lines along the $\IS^1$ fiber. They hence correspond to
different positions in 7 in the type I' picture. Different fractional D3,
D3' branes correspond to half D2 branes ending on the different NS branes,
located at different 7 positions. Our probe is a particular case of one
pair of fractional D3, D3' branes, with equal eigenvalue, specifying on
which NS brane the T dual half D2 brane is ending. Notice that the
Wilson line degree of freedom is to be considered a dynamical modulus in
the final field theory, and in that sense allows to continuously
interpolate between different choices of fractional D3, D3' branes.

Centering on this particular case of a single probe, we set without loss
of generality $n_0=n_0'=1$, and $n_i=n_i'=0$ for $i\neq 0$, and compute
the spectrum. The 33 spectrum is obtained by
the familiar projection, and leads to a $U(1)$ gauge theory with 8
supersymmetries. Expressing it in the language of 2d (4,4) susy
(namely working only with zero modes of the 4d fields), we get a (4,4)
$U(1)$ vector multiplet. Its four scalars parameterize the location of the
D3 brane (or the T dual D2 brane) in 2345. The 3'3' spectrum leads to a
similar spectrum. If 6 is non-compact, they are independent fields, but
with compact 6 they correspond to the same degrees of freedom and we get
only one copy of this spectrum. The 33' spectrum gives rise to one
hypermultiplet in the bifundamental representation. Since the D3 and D3'
branes have equal eigenvalue, it survives the orbifold projection. When
the direction 6 is taken compact, the bifundamental collapses to an
adjoint representation. The scalars in this field parameterize the
recombination of the two intersecting slags into a single smooth one, very
much like in \cite{mina3,mina}. Concretely, the deformed curve reads
\beqa
(z_1-i{\ov z}_2)(z_2-i {\ov z}_1) = \epsilon
\eeqa
and splits into two intersecting 2-cycles for $\epsilon=0$. Hence the full
2d spectrum on our probe is a (4,4) $U(1)$ vector multiplet and one (4,4)
adjoint, namely neutral, hypermultiplet. For $N$ overlapping probes, the
gauge symmetry is enhanced to $U(N)$.

The orientifold projection is easily analyzed. It maps D3 to D3' branes
and vice versa, reducing the spectrum (for compact 6) as follows. The
(4,4) $U(N)$ vector multiplet is projected down to a (0,4) $SO(N)$ vector
multiplet, and a (0,4) chiral multiplet (containing four real scalars and
four MW right-handed fermions) in the symmetric. The (4,4) hypermultiplet
is projected down to a (0,4) chiral multiplet in the two-index symmetric
representation, and a (0,4) `Fermi' multiplet (containing four MW left
fermions) in the antisymmetric representation. For $N=1$ we just get two
(0,4) chiral multiplets.

Notice that the $SO(N)$ theory we have described would suffer from 2d
gauge anomalies. However, one should recall that the full 2d theory also
contains states from strings stretched between the D2 and the D8 branes in
the type I' picture. They can be easily read out from this picture to
correspond to 32 MW left fermions in the fundamental of $SO(N)$, and
cancel the 2d anomaly. Alternatively, one can do the equivalent
computation in the type IIB orbifold picture, by introducing D7 branes at
angles, like in section 3, and computing the spectrum of strings between
the D3 and D7 branes. The cancellation of anomalies is a non-trivial check
that our procedure or reading the piece of the spectrum associated to the
intersection of D2, NS branes and O8 planes by using a T dual orbifold
construction is indeed consistent.

\medskip

It is time to ask for what physics the 2d LSM is describing. In particular 
if our computation has captured the LSM of the type I D1 brane in ALE
space, the Higgs moduli space of our theory should reproduce the 
background geometry felt by the type I D1 brane, a k-center Taub-Nut (or
ALE) space (different from the one in the type IIB orientifold
description). Unfortunately this seems not to be the case. 

In fact, there are several hints suggesting that our 2d field theory does
not really describe the propagation of a string in ALE space. In fact,
since the D3 brane probe in the type IIB orientifold is fractional, it
couples to at most one blow-up parameter, which can be shown to appear as
a FI term in the field theory as usual. This suggests that the D-brane
probe is sensitive to the geometry of just one TN center, not all $k$ of
them. Another piece of evidence comes from the actual 2d spectrum, which
is exactly that of a D1 brane probe in flat space. This suggests that the
moduli space of the theory is smooth, again suggesting the brane is
sensitive to just one TN center. Finally, this can be blamed, using the
T-dual type I' picture, to the fact that the D2 brane probe is sitting at
a definite position in 7 and the NS branes are located at different
positions in 7. Hence, the D2 brane can be made to coincide with at most
one NS brane. States that would become massless when the D2 brane
approaches other centers are generically massive. This fact is
geometrically manifest in the type I' picture, but not in the type IIB
orientifold setup, where it is however correctly recovered when one
computes the spectrum on the probe.

This should be understood as follows. By construction, the D1 brane probe
in type I must have full ALE geometry as its target space. Its worldvolume
theory has to contain parameters corresponding to all the blow-up modes. 
However, since the D1 string maps into a single D2 brane in type I' and
the single D2 maps only to a single fractional brane in our T-dual
picture, our calculation produces a field theory that is only sensitive to
a single blow-up. The issue is that in our calculation we have only
evaluated the massless sector. In order to probe the presence of the other
NS5 branes, we would have to include states whose mass is roughly ``twice
the distance'' to the other branes. Our perturbative orientifold is only
valid if the NS5 branes are equally spaced, so this distance is just
$R_7/k$. While one can in principle calculate the massive spectrum in the
orientifold as well, the problem is that the orientifold only captures the
near-core region of the full T-dual Taub-Nut geometry, so we are
neglecting states of mass $R_7$ anyway (which would come from strings
winding around the circle at infinity). The orientifold hence does not
encode the right spectrum at the massive level. So in order to get
the ADHM construction we were looking for, more refined tools to
evaluate the spectrum are necessary. We however hope that the T-dual
picture we have described is useful in further developments on this issue.

\section*{Acknowledgements}
We would like to thank Ofer Aharony, Gerardo Aldazabal, Ralph Blumenhagen, 
Mike Douglas, Sebasti\'an Franco, Luis Ib\'a\~nez, Boris K\"ors, Ra\'ul
Rabad\'an for very useful discussions, and especially Mina Aganagic for
many helpful questions and comments. A.~M.~U. thanks M.~Gonz\'alez for
kind encouragement and support. This work was partially supported by the
U.S. Department of Energy under contract \# DE-FC02-94ER40818. Y.H.H. is
also supported in part by the Presidential Fellowship of MIT.

\vspace{1.3cm}
\bibliographystyle{utphys}
\bibliography{linear}

\begingroup\raggedright\begin{thebibliography}{10}

\bibitem{hw}
A.~Hanany and E.~Witten, ``Type {IIB} superstrings, {BPS} monopoles, and
  three-dimensional gauge dynamics,'' {\em Nucl. Phys.} {\bf B492} (1997)
  152--190, \href{http://xxx.lanl.gov/abs/hep-th/9611230}{{\tt
  hep-th/9611230}}.

\bibitem{NSdual}
H.~Ooguri and C.~Vafa, ``Two-dimensional black hole and singularities of {CY}
  manifolds,'' {\em Nucl. Phys.} {\bf B463} (1996) 55--72,
  \href{http://xxx.lanl.gov/abs/hep-th/9511164}{{\tt hep-th/9511164}}.

\bibitem{ilka}
I.~Brunner and A.~Karch, ``Branes at orbifolds versus {Hanany Witten} in six-
  dimensions,'' {\em JHEP} {\bf 03} (1998) 003,
  \href{http://xxx.lanl.gov/abs/hep-th/9712143}{{\tt hep-th/9712143}}.

\bibitem{hanzaf}
A.~Hanany and A.~Zaffaroni, ``Branes and six-dimensional supersymmetric
  theories,'' {\em Nucl. Phys.} {\bf B529} (1998) 180,
  \href{http://xxx.lanl.gov/abs/hep-th/9712145}{{\tt hep-th/9712145}}.

\bibitem{fractional}
A.~Karch, D.~Lust, and D.~Smith, ``Equivalence of geometric engineering and
  hanany-witten via fractional branes,'' {\em Nucl. Phys.} {\bf B533} (1998)
  348, \href{http://xxx.lanl.gov/abs/hep-th/9803232}{{\tt hep-th/9803232}}.

\bibitem{pu}
J.~Park and A.~M. Uranga, ``A note on superconformal n = 2 theories and
  orientifolds,'' {\em Nucl. Phys.} {\bf B542} (1999) 139--156,
  \href{http://xxx.lanl.gov/abs/hep-th/9808161}{{\tt hep-th/9808161}}.

\bibitem{uranga}
A.~M. Uranga, ``Brane configurations for branes at conifolds,'' {\em JHEP} {\bf
  01} (1999) 022, \href{http://xxx.lanl.gov/abs/hep-th/9811004}{{\tt
  hep-th/9811004}}.

\bibitem{dandm}
K.~Dasgupta and S.~Mukhi, ``Brane constructions, conifolds and {M} theory,''
  \href{http://xxx.lanl.gov/abs/hep-th/9811139}{{\tt hep-th/9811139}}.

\bibitem{hanur}
A.~Hanany and A.~M. Uranga, ``Brane boxes and branes on singularities,'' {\em
  JHEP} {\bf 05} (1998) 013, \href{http://xxx.lanl.gov/abs/hep-th/9805139}{{\tt
  hep-th/9805139}}.

\bibitem{mina3}
M.~Aganagic, A.~Karch, D.~Lust, and A.~Miemiec, ``Mirror symmetries for brane
  configurations and branes at singularities,''
  \href{http://xxx.lanl.gov/abs/hep-th/9903093}{{\tt hep-th/9903093}}.

\bibitem{bs}
M.~Bianchi and A.~Sagnotti, ``On the systematics of open string theories,''
  {\em Phys. Lett.} {\bf B247} (1990) 517--524.

\bibitem{bs2}
M.~Bianchi and A.~Sagnotti, ``Twist symmetry and open string wilson lines,''
  {\em Nucl. Phys.} {\bf B361} (1991) 519--538.

\bibitem{gp}
E.~G. Gimon and J.~Polchinski, ``Consistency conditions for orientifolds and
  {D-Manifolds},'' {\em Phys. Rev.} {\bf D54} (1996) 1667--1676,
  \href{http://xxx.lanl.gov/abs/hep-th/9601038}{{\tt hep-th/9601038}}.

\bibitem{dp}
A.~Dabholkar and J.~Park, ``Strings on orientifolds,'' {\em Nucl. Phys.} {\bf
  B477} (1996) 701--714, \href{http://xxx.lanl.gov/abs/hep-th/9604178}{{\tt
  hep-th/9604178}}.

\bibitem{gj}
E.~G. Gimon and C.~V. Johnson, ``K3 orientifolds,'' {\em Nucl. Phys.} {\bf
  B477} (1996) 715--745, \href{http://xxx.lanl.gov/abs/hep-th/9604129}{{\tt
  hep-th/9604129}}.

\bibitem{intblum}
J.~D. Blum and K.~Intriligator, ``New phases of string theory and 6d {RG} fixed
  points via branes at orbifold singularities,'' {\em Nucl. Phys.} {\bf B506}
  (1997) 199, \href{http://xxx.lanl.gov/abs/hep-th/9705044}{{\tt
  hep-th/9705044}}.

\bibitem{hanzaf2}
A.~Hanany and A.~Zaffaroni, ``Monopoles in string theory,''
  \href{http://xxx.lanl.gov/abs/hep-th/9911113}{{\tt hep-th/9911113}}.

\bibitem{ralph}
R.~Blumenhagen, L.~Gorlich, and B.~Kors, ``Supersymmetric orientifolds in 6d
  with d-branes at angles,'' {\em Nucl. Phys.} {\bf B569} (2000) 209,
  \href{http://xxx.lanl.gov/abs/hep-th/9908130}{{\tt hep-th/9908130}}.

\bibitem{wade}
E.~Witten, ``Heterotic string conformal field theory and {A-D-E}
  singularities,'' \href{http://xxx.lanl.gov/abs/hep-th/9909229}{{\tt
  hep-th/9909229}}.

\bibitem{tensors}
J.~Polchinski, ``Tensors from {K3} orientifolds,'' {\em Phys. Rev.} {\bf D55}
  (1997) 6423--6428, \href{http://xxx.lanl.gov/abs/hep-th/9606165}{{\tt
  hep-th/9606165}}.

\bibitem{blpssw}
M.~Berkooz {\em et.~al.}, ``Anomalies, dualities, and topology of d=6 n=1
  superstring vacua,'' {\em Nucl. Phys.} {\bf B475} (1996) 115--148,
  \href{http://xxx.lanl.gov/abs/hep-th/9605184}{{\tt hep-th/9605184}}.

\bibitem{ralph4d}
R.~Blumenhagen, L.~Gorlich, and B.~Kors, ``Supersymmetric 4d orientifolds of
  type iia with d6-branes at angles,'' {\em JHEP} {\bf 01} (2000) 040,
  \href{http://xxx.lanl.gov/abs/hep-th/9912204}{{\tt hep-th/9912204}}.

\bibitem{fhs}
S.~Forste, G.~Honecker, and R.~Schreyer, ``Supersymmetric z(n) x z(m)
  orientifolds in 4d with d-branes at angles,'' {\em Nucl. Phys.} {\bf B593}
  (2001) 127--154, \href{http://xxx.lanl.gov/abs/hep-th/0008250}{{\tt
  hep-th/0008250}}.

\bibitem{gaberdiel}
B.~Craps and M.~R. Gaberdiel, ``Discrete torsion orbifolds and d-branes. ii,''
  \href{http://xxx.lanl.gov/abs/hep-th/0101143}{{\tt hep-th/0101143}}.

\bibitem{angelangle}
A.~M. Uranga, ``A new orientifold of c**2/z(n) and six-dimensional rg fixed
  points,'' {\em Nucl. Phys.} {\bf B577} (2000) 73,
  \href{http://xxx.lanl.gov/abs/hep-th/9910155}{{\tt hep-th/9910155}}.

\bibitem{ghm}
M.~B. Green, J.~A. Harvey, and G.~Moore, ``I-brane inflow and anomalous
  couplings on d-branes,'' {\em Class. Quant. Grav.} {\bf 14} (1997) 47--52,
  \href{http://xxx.lanl.gov/abs/hep-th/9605033}{{\tt hep-th/9605033}}.

\bibitem{scruse}
J.~Bogdan~Stefanski, ``Gravitational couplings of d-branes and o-planes,'' {\em
  Nucl. Phys.} {\bf B548} (1999) 275--290,
  \href{http://xxx.lanl.gov/abs/hep-th/9812088}{{\tt hep-th/9812088}}.

\bibitem{scruse2}
J.~F. Morales, C.~A. Scrucca, and M.~Serone, ``Anomalous couplings for d-branes
  and o-planes,'' {\em Nucl. Phys.} {\bf B552} (1999) 291--315,
  \href{http://xxx.lanl.gov/abs/hep-th/9812071}{{\tt hep-th/9812071}}.

\bibitem{cargese}
A.~Sagnotti, ``Open strings and their symmetry groups,''. Talk presented at the
  Cargese Summer Institute on Non- Perturbative Methods in Field Theory,
  Cargese, France, Jul 16-30, 1987.

\bibitem{ps}
G.~Pradisi and A.~Sagnotti, ``Open string orbifolds,'' {\em Phys. Lett.} {\bf
  B216} (1989) 59.

\bibitem{leroz}
R.~G. Leigh and M.~Rozali, ``Brane boxes, anomalies, bending and tadpoles,''
  {\em Phys. Rev.} {\bf D59} (1999) 026004,
  \href{http://xxx.lanl.gov/abs/hep-th/9807082}{{\tt hep-th/9807082}}.

\bibitem{amiandofer}
O.~Aharony and A.~Hanany, ``Branes, superpotentials and superconformal fixed
  points,'' {\em Nucl. Phys.} {\bf B504} (1997) 239,
  \href{http://xxx.lanl.gov/abs/hep-th/9704170}{{\tt hep-th/9704170}}.

\bibitem{review}
A.~Giveon and D.~Kutasov, ``Brane dynamics and gauge theory,'' {\em Rev. Mod.
  Phys.} {\bf 71} (1999) 983,
  \href{http://xxx.lanl.gov/abs/hep-th/9802067}{{\tt hep-th/9802067}}.

\bibitem{leighstrassler}
R.~G. Leigh and M.~J. Strassler, ``Exactly marginal operators and duality in
  four-dimensional n=1 supersymmetric gauge theory,'' {\em Nucl. Phys.} {\bf
  B447} (1995) 95--136, \href{http://xxx.lanl.gov/abs/hep-th/9503121}{{\tt
  hep-th/9503121}}.

\bibitem{wittenlift}
E.~Witten, ``Solutions of four-dimensional field theories via {M} theory,''
  {\em Nucl. Phys.} {\bf B500} (1997) 3,
  \href{http://xxx.lanl.gov/abs/hep-th/9703166}{{\tt hep-th/9703166}}.

\bibitem{phases}
E.~Witten, ``Phases of {N=2} theories in two-dimensions,'' {\em Nucl. Phys.}
  {\bf B403} (1993) 159--222,
  \href{http://xxx.lanl.gov/abs/hep-th/9301042}{{\tt hep-th/9301042}}.

\bibitem{edeva}
E.~Silverstein and E.~Witten, ``Global {U(1) R} symmetry and conformal
  invariance of (0,2) models,'' {\em Phys. Lett.} {\bf B328} (1994) 307--311,
  \href{http://xxx.lanl.gov/abs/hep-th/9403054}{{\tt hep-th/9403054}}.

\bibitem{polchinski}
J.~Polchinski and E.~Witten, ``Evidence for heterotic - type {I} string
  duality,'' {\em Nucl. Phys.} {\bf B460} (1996) 525--540,
  \href{http://xxx.lanl.gov/abs/hep-th/9510169}{{\tt hep-th/9510169}}.

\bibitem{comments}
E.~Witten, ``Some comments on string dynamics,''
  \href{http://xxx.lanl.gov/abs/hep-th/9507121}{{\tt hep-th/9507121}}.

\bibitem{michaofer}
O.~Aharony and M.~Berkooz, ``Ir dynamics of d = 2, n = (4,4) gauge theories and
  dlcq of 'little string theories','' {\em JHEP} {\bf 10} (1999) 030,
  \href{http://xxx.lanl.gov/abs/hep-th/9909101}{{\tt hep-th/9909101}}.

\bibitem{pru}
J.~Park, R.~Rabadan, and A.~M. Uranga, ``N = 1 type iia brane configurations,
  chirality and t- duality,'' {\em Nucl. Phys.} {\bf B570} (2000) 3--37,
  \href{http://xxx.lanl.gov/abs/hep-th/9907074}{{\tt hep-th/9907074}}.

\bibitem{mina}
M.~Aganagic and C.~Vafa, ``Mirror symmetry, d-branes and counting holomorphic
  discs,'' \href{http://xxx.lanl.gov/abs/hep-th/0012041}{{\tt hep-th/0012041}}.

\end{thebibliography}\endgroup
\end{document}